\documentstyle{mn}

\title[Intermittency in the Galaxy Distribution]{The Origin of Spatial Intermittency in the Galaxy Distribution}

\author[Peter Coles]{Peter Coles\\
School of Physics \& Astronomy, University of Nottingham,
University Park, Nottingham, NG7 2RD, United Kingdom\\ }

\begin{document}

\maketitle

\begin{abstract}
The dynamical equations describing the evolution of a
self-gravitating fluid can be rewritten in the form of a
Schr\"{o}dinger equation coupled to a Poisson equation determining
the gravitational potential. This approach has a number of
interesting features, many of which were pointed out in a seminal
paper by Widrow \& Kaiser (1993). In particular we show that this
approach yields an elegant reformulation of an idea due to Jones
(1999) concerning the origin of lognormal intermittency in the
galaxy distribution.
\end{abstract}


\section{Introduction}

Galaxy redshift surveys such as the 2dF Galaxy redshift survey and
the Sloan Digital Sky Survey are poised to yield a copious harvest
of statistical information about the distribution and dynamical
properties of the large-scale structure of the Universe. At the
same time relatively few statistical properties of the structure
revealed by these observational programs are understood in
quantitative detail using analytic methods.

One particular facet that has received some attention over the
years has been the property that the one-point probability
distribution of density fluctuations $p(\rho)$ appears to have a
roughly lognormal form, i.e. $\log \rho$ has a roughly normal
distribution (Coles \& Jones 1991). It is now established that
this property has an interesting connection with the scaling
properties of moments of the probability of the distribution.
Taking a generic random variable $X$, such that the distribution
of $X$ within cells of side $L$ is denoted $p(X;L)$, then the
$q$-th moment at a given value of $L$ is said to display scaling
if
\begin{equation}
\langle X^q \rangle_L = \sum p(X;L)X^q \propto L^{\mu(q)}.
\end{equation}
This means that different powers $q$ of the density field vary as
a different power of the coarse-graining scale $L$. The function
$\mu(q)$ is called the {\em intermittency exponent}, and it can be
extracted from observations. Jones, Coles \& Martinez (1992)
showed that observations suggest a roughly quadratic dependence of
$\mu(q)$ upon $q$ and that this is related to the underlying
near-lognormal form of the density fluctuations. A set displaying
the form (1) is usually termed a multifractal; see Paladin \&
Vulpiani (1987) for general discussion.

The term {\em intermittency} was coined to described properties of
stochastic processes described by high skewed probability
distributions with very slowly decaying tails. The lognormal is a
prime example, used in a pioneering paper by Kolmogorov (1962). In
the time domain it refers to processes involving extended
quiescent periods interrupted by bursts of intense activity. In
the spatial domain, intermittent processes are ones in which
isolated regions of high density are separated by large voids.
Although in a qualitative sense the application of the concept of
intermittency to large-scale structure seems plausible, a
quantitative description of how it arises is not easy to obtain.
Drawing on ideas discussed by Zel'dovich et al. (1985, 1987),
Jones (1999) suggested an analytical model for the cosmological
context. It is the purpose of this paper to describe an
alternative formulation of the Jones (1999) model and
substantially strengthens the physical understanding of this
model. This also provides an opportunity to advocate the wider use
of an alternative formulation of the gravitational instability
scenario discussed by Widrow \& Kaiser (1993).

\section{The Fluid Approach to Structure Formation}

\subsection{Basics}
In order to understand how the intermittent form of
large-scale structure arises, it is best to begin with the
standard fluid-based approach to structure growth. In the standard
treatment of the Jeans Instability one begins with the dynamical
equations governing the behaviour of a self-gravitating fluid.
These are: the {\em Euler equation}
\begin{equation}
{\partial ({\bf v})\over \partial t} + ({\bf v}\cdot{\bf
\nabla}){\bf v}  + {1\over \rho}{\bf \nabla} p + {\bf
\nabla}\phi=0~; \label{eq:Euler1}
\end{equation}
the {\em continuity equation} \begin{equation}
{\partial\rho\over
\partial t} +  +  {\bf \nabla} (\rho{\bf v}) =
0 \label{eq:continuity1}
\end{equation}
expressing the conservation of matter; and the {\em Poisson
equation}
\begin{equation} {\bf
\nabla}^2\phi = 4\pi G \rho, \label{eq:Poisson1}
\end{equation}
describing Newtonian gravity.

\subsection{The Cosmological Setting}

If the length scale of the perturbations is smaller than the
effective cosmological horizon $d_H=c/H_0$, a Newtonian treatment
of cosmic structure formation is still expected to be valid in
expanding world models. In an expanding cosmological background,
the Newtonian equations governing the motion of gravitating
particles can be written in terms of
\begin{equation}
 {\bf x} \equiv {\bf r} / a(t)\end{equation} (the comoving spatial
coordinate, which is fixed for observers moving with the Hubble
expansion),
\begin{equation} {\bf v} \equiv \dot {{\bf r}} - H {\bf r} = a\dot
{{\bf x}}\end{equation} (the peculiar velocity field, representing
departures of the matter motion from pure Hubble expansion), $\rho
({\bf x}, t)$ (the matter density), and $\phi ({\bf x} , t)$ (the
peculiar Newtonian gravitational potential, i.e. the fluctuations
in potential with respect to the homogeneous background)
determined by the Poisson equation in the form
\begin{equation}
{\bf \nabla_x}^2\phi = 4\pi G a^2(\rho - \rho_0) = 4\pi
Ga^2\rho_0\delta. \label{eq:Poisson}
\end{equation}
In this equation and the following the suffix on $\nabla_x$
indicates derivatives with respect to the new comoving
coordinates. Here $\rho_0$ is the mean background density, and
\begin{equation}
\delta \equiv \frac{\rho-\rho_0}{\rho_0}
\end{equation}
is the {\em density contrast}. Using these variables the Euler
equation becomes
\begin{equation}
{\partial (a{\bf v})\over \partial t} + ({\bf v}\cdot{\bf
\nabla_x}){\bf v} = - {1\over \rho}{\bf \nabla_x} p - {\bf
\nabla_x}\phi~. \label{eq:Euler}
\end{equation}
The first term on the right-hand-side of equation (\ref{eq:Euler})
arises from pressure gradients, and is neglected in dust-dominated
cosmologies. Pressure effects may nevertheless be important in the
the (collisional) baryonic component of the mass distribution when
nonlinear structures eventually form. The second term on the
right-hand side of equation (\ref{eq:Euler}) is the peculiar
gravitational force, which can be written in terms of ${\bf g} =
-{\bf \nabla_x}\phi/a$, the peculiar gravitational acceleration of
the fluid element. If the velocity flow is irrotational, ${\bf v}$
can be rewritten in terms of a velocity potential $\phi_v$:
\begin{equation}{\bf v} = - {\bf
\nabla_x} \phi_v/a. \end{equation} This is expected to be the case
in the cosmological setting because (a) there are no sources of
vorticity in these equations and (b) vortical perturbation modes
decay with the expansion. Next we have the revised continuity
equation:
\begin{equation}
{\partial\rho\over \partial t} + 3H\rho + {1\over a} {\bf
\nabla_x} (\rho{\bf v}) = 0, \label{eq:continuity}
\end{equation}

\subsection{The Nonlinear regime}

The strength of the fluid approach is that it furnishes a
relatively simple perturbative treatment for the early stages of
structure formation, wherein the density fluctuations are small.
Perturbation theory fails when nonlinearities develop, and
unfortunately this it is in the nonlinear regime where scaling and
intermittency arise. It is important to stress, however, that the
fluid treatment is intrinsically approximate anyway. A fuller
treatment of the problem requires a solution of the Boltzmann
equation for the full phase-space distribution of the system
$f({\bf x}, {\bf v}, t)$ coupled to the Poisson equation
(\ref{eq:Poisson1}) that determines the gravitational potential.
In cases where the matter component is collisionless, the
Boltzmann equation takes the form of a Vlasov equation:
\begin{equation}
{\partial f \over \partial t}= \sum_{i=1}^{3} \left({\partial
\phi\over
\partial x_i}{\partial f \over \partial v_i} - v_i {\partial f \over \partial
x_i}\right).
\end{equation}
The fluid approach  can only describe cold material where the
velocity dispersion of particles is negligible. But even if the
dark matter is cold, there may be hot components of baryonic
material whose behaviour needs also to be understood. Moreover,
the fluid approach assumes the existence of a single fluid
velocity at every spatial position. It therefore fails when orbits
cross and multi-streaming generates a range of particle velocities
through a given point. While the existence and location of the
caustics that occur and shell crossing can be determined using the
Zel'dovich (1970) approximation (Coles, Melott \& Shandarin 1993),
the dynamics and distribution of matter after shell-crossing is
not described in this approach. Attempts to understand properties
of non-linear structure using the fluid model therefore resort to
further approximations (Sahni \& Coles 1995) to extend the
approach beyond shell-crossing. Even the numerical N-body methods
that many regard as the ultimate approach to this kind of problem
are also approximate, corresponding to a particular type of Monte
Carlo integration of the Vlasov equation.

\subsection{The Jones Model}
It is within the overall framework of the fluid model that Jones
(1999) sought to understand the observed intermittency of the
large-scale structure of the Universe. Using the velocity
potential introduced above, he first introduces an effective
Bernoulli equation for the flow:
\begin{equation}
{\partial \phi_v \over \partial t} - {(\nabla \phi_v)^2\over
2a^2}=\phi,
\end{equation}
where $\phi$ is the actual gravitational potential. This equation
neglects terms involving pressure gradients as mentioned above. To
cope with shell-crossing events, Jones (1999) introduces an
artificial viscosity $\nu$ by adding a term to the right-hand-side
of this equation:
\begin{equation}
{\partial \phi_v \over \partial t} - {(\nabla \phi_v)^2\over
2a^2}=\phi+{\nu\over a^2} \nabla^2\phi_v.
\end{equation}
The viscosity is introduced to prevent the particle flow from
entering the multi-stream region by causing particles to stick
together at shell-crossing. This is also used in an approach
called the adhesion approximation (Gurbatov, Saichev \& Shandarin
1989). Using the Hopf-Cole transformation $\phi_v=-2\nu\log
\varphi$ and defining a scaled gravitational potential via
$\phi=2\nu\epsilon$ we can write the Bernoulli equation as
\begin{equation}
{\partial \varphi \over \partial t} = \nu \nabla^2 \varphi +
\epsilon({\bf x}) \varphi. \label{eq:heat}
\end{equation}
This is called the random heat equation, because of the existence
of the spatially-fluctuating potential term $\epsilon({\bf x})$.
The gravitational potential changes very slowly even during
nonlinear evolution (Brainerd, Scherrer \& Villumsen 1993; Bagla
\& Padmanabhan 1994), so Jones (1999) assumes that it can be taken
as constant and to be Gaussian distributed. An approximate scaling
solution to (\ref{eq:heat}) can then be found using a path
integral adapted from that normally used in quantum physics
(Feynman \& Hibbs 1965); see below for more details. In this
approximation, the function $\varphi$ has a lognormal distribution
(Coles \& Jones 1991). We refer the reader to Jones (1999) for
details; see also Zel'dovich et al. (1985, 1987).

This model is one of the few attempts that have been made to
understand the non-linear behaviour of the matter distribution
using analytic methods. Although not rigorous it surely captures
the essential factors involved. It does, however, suffer from a
number of shortcomings. First, the approach does not follow
material beyond the shell-crossing stage. Second, the viscosity
$\nu$ that is needed does not have properties that are very
realistic physically:  it can depend neither on the density $\rho$
nor the position ${\bf x}$. Moreover, in the final analysis Jones
(1999) takes the limit $\nu\rightarrow 0$, so it cancels out
anyway. One is tempted to speculate that its introduction may be
unnecessary. Third, the function $\varphi({\bf x},t)$ that emerges
from equation (\ref{eq:heat}) is not the desired density
$\rho({\bf x}, t)$ nor does it bear a straightforward relation to
the density. Finally, it is not clear how the motion of a
collisional baryonic component can be modelled within this
framework.

\section{Wave Mechanics and Structure Formation}
A novel approach to the study of collisionless matter was
suggested by Widrow \& Kaiser (1993). They advocate the
replacement of the Euler and continuity equations by a version of
the Schr\"{o}dinger equation
\begin{equation}
i\hbar {\partial \psi\over \partial t}= - {\hbar^2 \over 2m}
\nabla^2\psi + m \phi({\bf x})\psi.\label{eq:schrod}
\end{equation}
The equivalence between this and the fluid approaches has been
known for some time; see Spiegel (1980) for historical comments.
Originally the interest was to find a fluid interpretation of
quantum mechanical effects, but in this context we shall use it to
describe an entirely classical system. In this spirit, the
constant $\hbar$ is taken to be an adjustable parameter that
controls the spatial resolution $\lambda$ through a de Broglie
relation $\lambda=\hbar/mv$. To accommodate gravity we need to
couple  equation (\ref{eq:schrod}) to the Poisson equation in the
form
\begin{equation}
\nabla^2\phi=4\pi G\psi \psi^*.
\end{equation}
It clearly follows that $\rho=|\psi|^2$; the wavefunction
$\psi({\bf x}, t)$ evidently complex.  Moreover, if the matter
distribution is initially cold then one can further construct a
wavefunction that also encodes the velocity part of phase space in
its argument through the ansatz
\begin{equation}
\psi({\bf x})=\sqrt{\rho({\bf x})} \exp [i\theta({\bf x}/\hbar)],
\end{equation}
where $\nabla \theta({\bf x})={\bf p}({\bf x})$, the local
`momentum field'. This formalism thus yields an elegant
description of both the density and velocity fields in a single
function.

This formulation provides a useful complementary approach to many
techniques, including $N$-body methods. It also provides a new
light with which to study the Jones (1999) model. Widrow \& Kaiser
(1993) show using theoretical arguments and numerical simulations
that this system allows accurate numerical evolution of the system
beyond shell-crossing, so it does not have the ad hoc construction
needed by the Jones (1999) model to remedy this.

Second, no artificial viscosity is required. Equations
(\ref{eq:heat}) and (\ref{eq:schrod}) are of the same form, apart
from minor subtleties like the use of complex time coordinates.
The potential term is easily understood in (\ref{eq:schrod}), and
the wavefunction $\psi$ now has a straightforward relationship to
$\rho$. The upshot of this is that if one adopts the approximation
of constant gravitational potential one can use the same path
integral approach as described by Jones (1999). In a nutshell,
given some initial wavefunction $\psi({\bf x'}, t')$ one can
determine the wavefunction at a subsequent time $\psi({\bf x}, t)$
using
\begin{equation}
\psi({\bf x}, t)=\int K({\bf x}, t; {\bf x'}, t') \psi ({\bf x'},
t') d^3x',
\end{equation}
where the function $K({\bf x}, t; {\bf x'}, t')$ involves a sum
over all paths $\Gamma$ connecting the initial and final states
with $t>t'$:
\begin{equation}
K({\bf x}, t; {\bf x'}, t')=\int {\cal D}\Gamma \exp
[iS(\Gamma)/\hbar]\label{eq:Green},
\end{equation}
where ${\cal D}$ is an appropriate measure on the set of classical
space-time trajectories. For a particle moving in a potential
$V({\bf x}, t)=m\phi({\bf x}, t)$ the action $S$ for a given path
$\Gamma$ is given by
\begin{equation}
S(\Gamma)=\int_{\Gamma} \left[ {1\over 2} m \dot{x}^2 - m\phi({\bf
x}) \right] dt.
\end{equation}
Note the presence of the Gaussian field in equation (21) and hence
in the exponential of the integrand  on the right-hand-side of
equation (\ref{eq:Green}). To get an approximate solution to this
system we can follow the same reasoning as Zeldovich et al. (1985,
1987) and Jones (1999), ignoring time-varying terms, using the
Gaussian properties and counting the dominant contributions to the
path integral to deduce that the integral produces a solution of
lognormal form. This part of the argument is identical to that
advanced by Jones, except that the solution is for $\psi$ rather
than $\varphi$ and since $\rho$ is $|\psi^2|$ then one directly
obtains a lognormal distribution for the desired density
$\rho({\bf x}, t)$.

It should be stressed that, although the present approach clearly
provides a more elegant formulation of the problem, the deduction
of lognormality remains approximate; the lognormal is not the
exact solution to the system to either Jones' equation (15) or the
present equation (16). How accurately this approximate form
applies is open to doubt and will have to be checked by full
numerical solutions. Interestingly, however, it is known to apply
quite accurately in quantum systems such as disordered mesoscopic
electron configurations (Janssen 1998). As mentioned above, the
Schrodinger approach yields a wavefunction $\psi$ which is
directly related to the particle density $\rho$ via
$\rho=|\psi|^2$. Such quantum systems also display lognormal
scaling for properties such as the conductance, which depends on
$|\psi|^4$. It is a property of the lognormal distribution that if
a random variable $X$ is lognormal, then so is $X^n$. In such
systems the role of the gravitational potential $\phi$ is played
by a potential that describes the disorder of a solid, perhaps
caused by the presence of defects. Such systems display {\em
localisation} at low temperature which is similar in some ways to
the original idea of Anderson location (Anderson 1958). The
formation of strongly non-linear structures by gravity is thus
directly analogous to the generation of localised wavefunctions in
condensed matter systems.

Finally, and perhaps most promisingly for future work, the
equation (\ref{eq:schrod}) offers a relatively straightforward way
of modelling the behaviour of collisional material. The addition
to the potential of a term of the form $\alpha |\psi|^2$ (with
$\alpha$ an appropriately-chosen constant), converts the original
equation (\ref{eq:schrod}) into a nonlinear Schr\"{o}dinger
equation:
\begin{equation}
i\hbar {\partial \psi\over \partial t}= - {\hbar^2 \over 2m}
\nabla^2\psi + m \phi({\bf x})\psi + \alpha |\psi|^2
\psi\label{eq:schrodnl}
\end{equation}
(Sulem \& Sulem 1999). This equation is now equivalent to those
that describe the flow of a barotropic fluid; see Spiegel (1980).
This system can therefore be used to model pressure effects, which
are otherwise only handled effectively using numerical methods
such as smoothed-particle hydrodynamical approximations (e.g.
Monaghan 1992). In the context of quantum systems, the nonlinear
term is used to describe the formation of Bose-Einstein
condensates (e.g. Choi \& Niu 1999 and reference therein).

\section{Discussion}

In this short paper I have sketched out an approach to the study
evolving cosmological density fluctuations that relies on a
transformation of the Vlasov-Poisson system into a
Schr\"{o}dinger-Poisson system. The transformation is not a new
idea, but despite the efforts of Widrow \& Kaiser (1993) it does
not seem to be well known in the astronomical community. The
immediate advantage of this new formalism is that it yields a
rather more convincing approach to understanding the origin of
spatial intermittency and approximate lognormality in the galaxy
distribution than that offered by Jones (1999). It also makes a
connection in the underlying physics with other systems that
display similar phenomena.

On the other hand, one must be aware of the approximations also
inherent in the present approach. The Schr\"{o}dinger equation is
not exact, and its usefulness as an approximate tool is restricted
by a number of conditions outlined by Widrow \& Kaiser (1993); see
also Spiegel (1980). Furthermore, the lognormal solution of the
system is a further approximation and may not be valid especially
in the strongly-fluctuating limit. Although it neatly bypasses
some of the problems inherent in the Jones (1999) model, the
nonlinear wave equation is by no means easy to solve in general
situations. Numerical methods will still have to be employed to
understanding other aspects of the evolution of cosmic structure
within this framework as indeed they are in other branches of
physics.

One particular issue worth exploring using this approach is to
understand the limits of the approach in strongly non-linear
situations. As it stands, the justification for the lognormal
approximation arises from the weakly non-linear behaviour of
collisionless matter moving in an almost constant potential field.
Taking into account the expansion of the Universe, the changing
gravitational potential, and the possible effects of matter
pressure within in the action formalism described in Section 3 may
well reveal that a different form of hierarchical scaling pertains
in the strongly non-linear regime. One aspect of this is that the
hierarchy of correlation functions that describe a lognormal
distribution display Kirkwood (1935) scaling, while it appears
from numerical $N$-body simulations that cosmological fluctuations
display a different hierarchical form. For a discussion of the
relationship between lognormal  and hierarchical scaling, see
Coles \& Frenk (1991).

Apart from the particular topic of intermittency, this paper is
also intended to stimulate interest in this general approach. Many
techniques exist for studying the wave mechanics of disordered
systems, such as the renormalization group and path-integral
methods, few of which are used by cosmologists working in this
area. It is to be hoped that the introduction of some of these
methods may allow better physical insights into the behaviour of
non-linear structure formation than can be found using brute-force
$N$-body techniques.


\begin{thebibliography}{}
\bibitem{} Anderson P.W., 1958, Phys. Rev., 109, 1492
\bibitem{} Bagla J.S., Padmanabhan T., 1994, MNRAS, 266, 227
\bibitem{} Brainerd T.G., Scherrer R.J., Villumsen J.V., 1993,
ApJ, 418, 570
\bibitem{} Choi D.-I., Niu Q., 1999, Phys. Rev. Lett., 82, 2022
\bibitem{} Coles P., Frenk C.S., 1991, MNRAS, 253, 727
\bibitem{} Coles P., Jones B.J.T., 1991, MNRAS, 248, 1
\bibitem{} Coles P., Melott A.L. \& Shandarin S.F., 1993, MNRAS,
260, 765
\bibitem{} Feynman R.P., Hibbs A.R., 1965, Quantum
Mechanics and Path Integrals. McGraw Hill, New York
\bibitem{} Gurbatov S.N., Saichev A.I., Shandarin S.F., 1989,
MNRAS, 236, 385
\bibitem{} Janssen M., 1998, Physics Reports, 295, 1
\bibitem{} Jones B.J.T., 1999, MNRAS, 307, 376
\bibitem{} Jones B.J.T., Coles P., Martinez V.J., 1992, MNRAS,
259, 246
\bibitem{} Kirkwood J.C., 1935, J. Chem. Phys., 3, 300
\bibitem{} Kolmogorov A.N., 1962, J. Fluid Mech., 13, 82
\bibitem{} Monaghan J.J., 1992, ARA\& A, 30, 543
\bibitem{} Paladin G., Vulpiani A., 1987, Physics Reports, 156,
147
\bibitem{} Sahni V., Coles P., 1995, Physics Reports, 262, 1
\bibitem{} Shandarin S.F., Zel'dovich Ya. B., 1989, Rev. Mod.
Phys., 61, 185
\bibitem{} Spiegel E.A., 1980, Physica, 1D, 236
\bibitem{} Sulem C., Sulem P.-L., 1999, The Nonlinear Schrodinger
Equation. Self focusing and Wave Collapse. Springer, New York
\bibitem{} Widrow L.M., Kaiser N., 1993, ApJ, 416, L71
\bibitem{} Zel'dovich Ya.B., 1970, A\& A, 5, 84
\bibitem{} Zel'dovich Ya. B., Molchanov S.A., Ruzmaikin A.A.,
Sokolov D.D., 1985, Sov. Phys. JETP, 62, 1188
\bibitem{} Zel'dovich Ya. B., Molchanov S.A., Ruzmaikin A.A.,
Sokolov D.D., 1987, Sov. Phys. Usp., 152, 3
\end{thebibliography}
\end{document}